\documentclass{iau}
\usepackage{graphicx}

\title[LACEwING: A New Moving Group Code] 
{LACEwING: Lessons from a New Moving Group Code}

\author[A. R. Riedel]   
{A. R. Riedel$^1$}

\affiliation{$^1$The College of Staten Island\\email: {\tt adric.riedel@csi.cuny.edu}\\
CUNY-The College of Staten Island\\
Hunter College\\
The American Museum of Natural History}

\pubyear{2015}
\volume{314}  
\pagerange{119--126}
\jname{Young Stars \& Planets Near the Sun}
\editors{J. H. Kastner, B. Stelzer, \& S. A. Metchev, eds.}
\begin{document}

\maketitle

\begin{abstract}
With the recent accelerating rate of discoveries in the field of nearby young stars, the ability to identify new nearby young stars is as important as ever, and membership identification codes will continue to perform a vital role in scientific research. In the process of creating a new moving group membership identification code - LocAting Constituent mEmbers In Nearby Groups (LACEwING) - we have compiled a few pointers relevant to astronomers trying to use codes like LACEwING to locate young stars.
\keywords{stars:low-mass,stars:pre-main-sequence,galaxy:open clusters and associations}
\end{abstract}

\firstsection 
\section{Introduction}

One of the most important methods used in characterizing young stars, and the only method that can establish membership in specific nearby young moving groups, is their kinematics. Kinematics exploits the property that young stars (at ages less than a billion years) are still tracing the space motion of the gas from which they formed. Several membership identification codes are available publically: BANYAN \cite[(Malo et al. 2013)]{Malo2013}, BANYAN II \cite[(Gagne et al. 2014)]{Gagne2014}; a convergence code \cite[(Rodriguez et al. 2013)]{Rodriguez2013}; and (soon), LACEwING \cite[(Riedel et al. in prep)]{Riedel2015}\footnote{See also https://github.com/ariedel/lacewing}. 

These codes use the six basic kinematic elements (RA, DEC, $\pi$, $\mu_{RA}$, $\mu_{DEC}$, and RV) to predict probabilities of membership in nearby young moving groups, and can handle incomplete data. It is, however, important to know how to use and interpret the results of LACEwING (and codes like it) to obtain the most accurate results.

\section{Overview of LACEwING}
The LACEwING code computes membership probabilities in 10 moving groups ($\epsilon$ Chameleon, TW Hydra, $\beta$ Pic, Octans, Tucana-Horologium, Columba, Argus, AB Doradus, Hercules-Lyra, and Ursa Major) and 4 open clusters ($\eta$ Chameleon, The Pleiades, Coma Berenices, and The Hyades). The groups are represented as freely oriented triaxial ellipsoids, which are computed from new lists of bona-fide (high confidence) members taken from existing lists (e.g. \cite[Malo et al. 2013]{Malo2013}), and filtered using epicyclic tracebacks to limit membership to only the stars that were plausibly within the group's boundaries at the time of formation.

\begin{figure}[h]
\begin{center}
\includegraphics[width=2.5in]{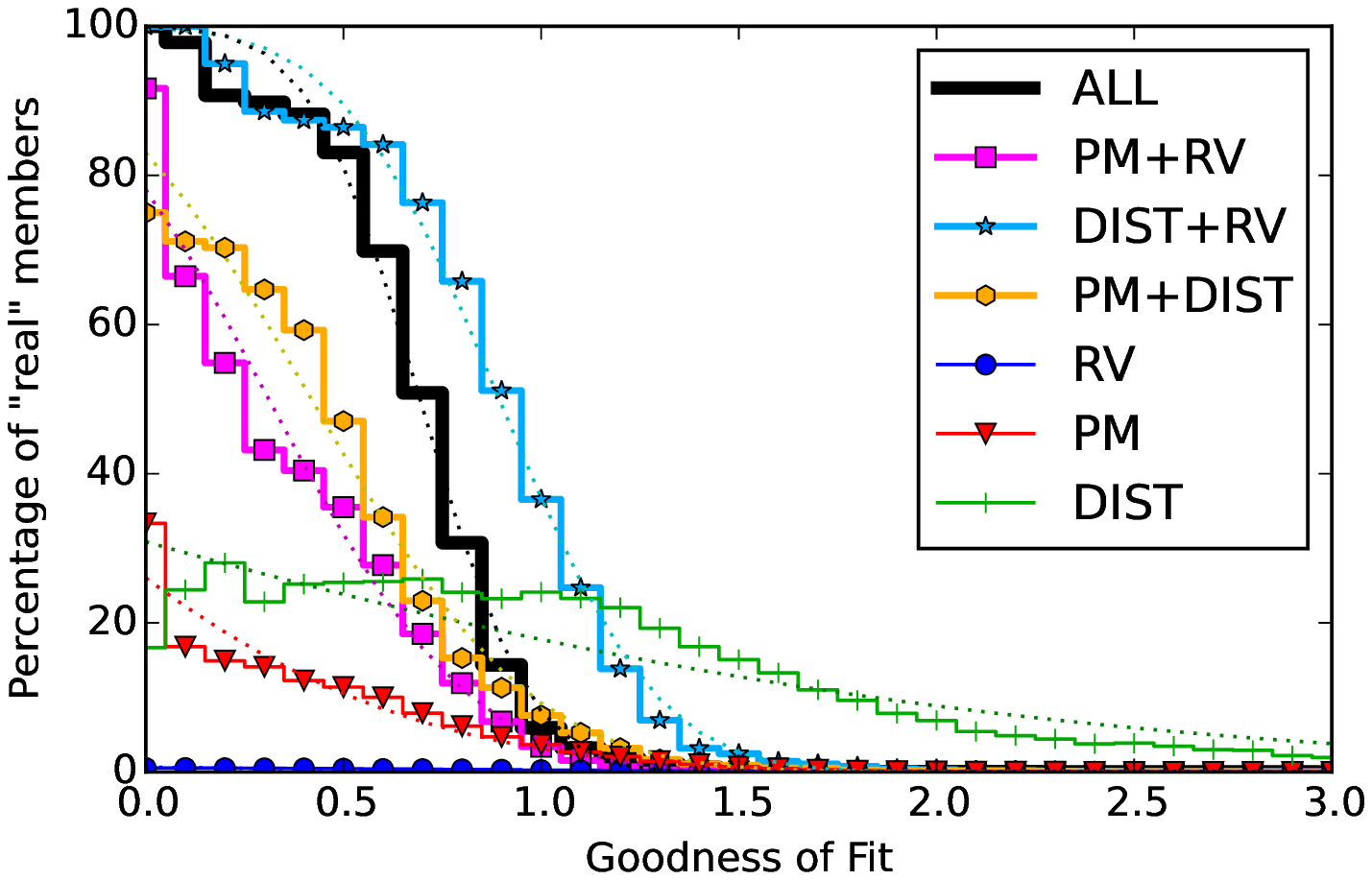} 
\includegraphics[width=2.5in]{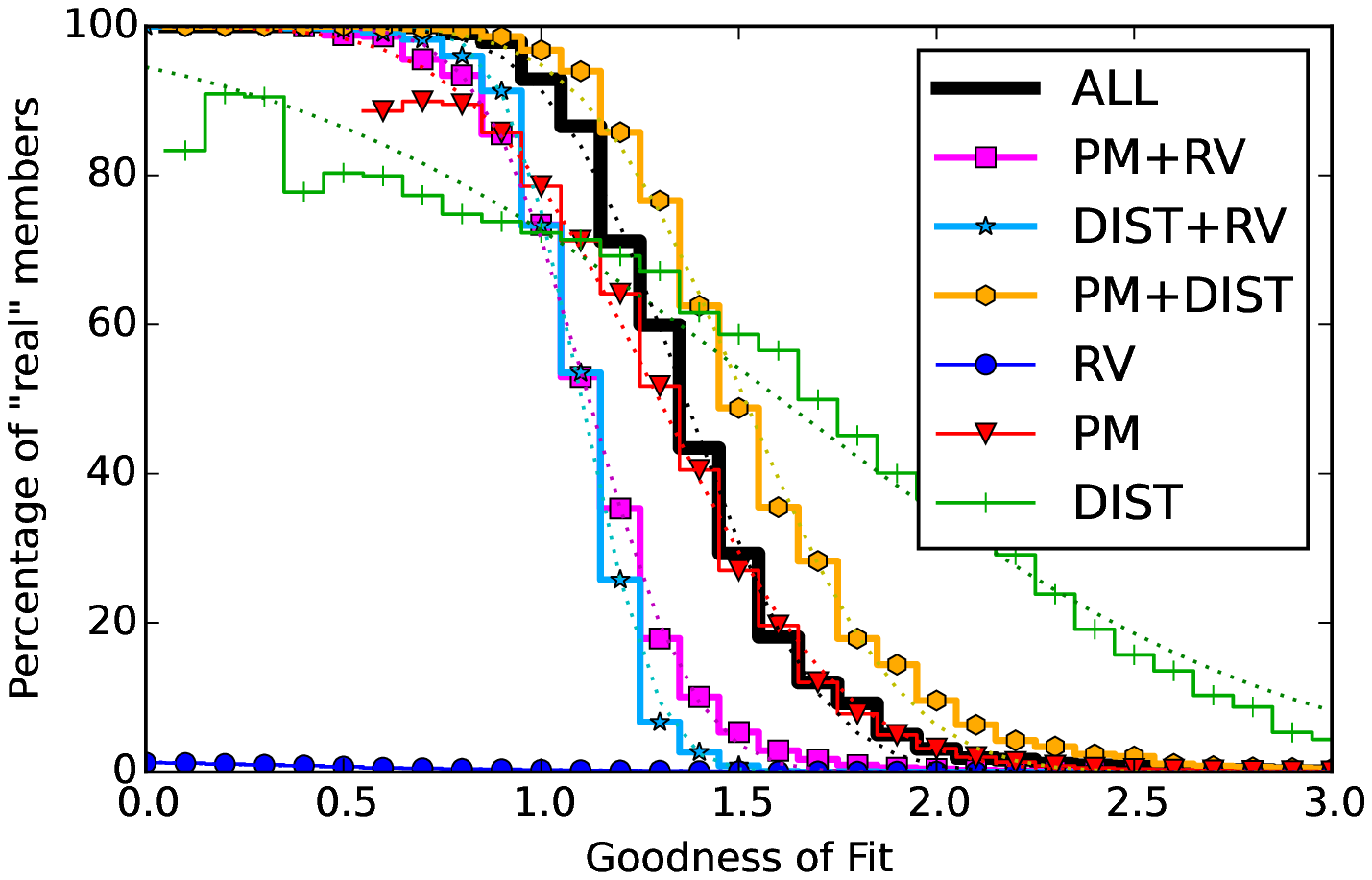} 
 \caption{Histograms of matches to the simulated $\beta$ Pictoris Moving Group (left) and Coma Berenices open cluster (right), showing the percentage of stars in each bin that are genuine members of the group/cluster. Curves fit to these histograms (dotted lines) are used to translate the LACEwING code's goodness-of-fit parameters to actual membership probabilities.
   \label{fig:beta_pic}}
\end{center}
\end{figure}

The LACEwING code matches stars to groups by computing up to four metrics of membership, matching the proper motion, distance, radial velocity, and spatial location of the star to those estimated for a group member at the same RA and DEC. Those metrics are combined into a single numerical goodness-of-fit value. The goodness-of-fit values are transformed into membership probabilities by creating a large simulation (using 32 million stars; \cite[Riedel et al. submitted]{Riedel2015}) of the Solar Neighborhood, which is then binned by goodness-of-fit value relative to a group. This is shown for $\beta$ Pic and Coma Ber in Fig.\,\ref{fig:beta_pic}, where all seven different possible combinations of data (RA and DEC are assumed to be known) are given as different colors. Curves are fit to those distributions to provide membership probabilities.

\section{Lessons from LACEwING}

\subsection{\bf Groups Are Not Equally Easy to Identify}

\begin{figure}[h]
\begin{center}
\includegraphics[width=5in]{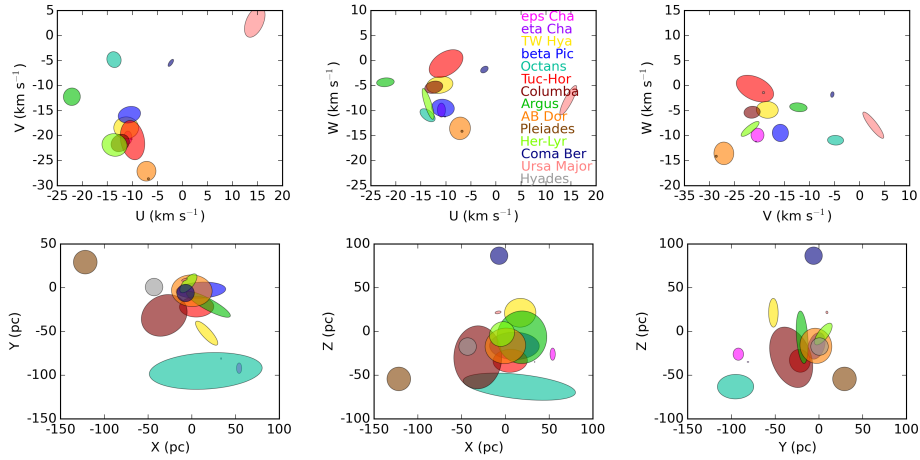} 
 \caption{2D projections of the freely oriented triaxial ellipsoids that constitute LACEwING's kinematic models (in UVW motion and XYZ spatial position) of the 14 nearby young moving groups and open clusters. As shown in these figures, some groups have overlapping UVW space velocities or XYZ space positions. This can make it difficult to determine to which group individual stars belong.
   \label{fig:ellipses}}
\end{center}
\end{figure}

Some groups (like the Coma Berenices open cluster) have very distinctive UVW space velocities and space positions (Fig.\,\ref{fig:ellipses}), which translate into uniquely identifying proper motion and radial velocity estimates for any given star. 

As an example, the large simulation of stars in Fig.\,\ref{fig:beta_pic} shows us that for the case in which all kinematic elements are available (RA, DEC, plus proper motion, distance, and rv), a goodness-of-fit value of 0.75 relative to $\beta$ Pic yields a roughly 50\% probability that the object is a member of $\beta$ Pic. The stars matched to Coma Ber show that, for the equivalent case using all available data, a goodness-of-fit of 0.75 yields a nearly 100\% probability that the object is a member of Coma Ber.

\subsection{\bf Additional Kinematic Data Vastly Improves Membership Probabilities} 
Fig.\,\ref{fig:beta_pic} shows the seven different data availability scenarios LACEwING expects: RA and DEC plus all combinations of proper motion, radial velocity, and distance. For $\beta$ Pic, the proper motion method (red dashed line) demonstrates that even in the best possible case of a goodness-of-fit = 0, the probability of membership is only 35\%, i.e. 65\% of those perfect matches to $\beta$ Pic are actually (simulated) members of other moving groups and the generic field population. Adding a radial velocity (pink triangles) increases the best-case probability to 80\%/20\% contamination. 

\begin{figure}[h]
\begin{center}
\includegraphics[width=2.5in]{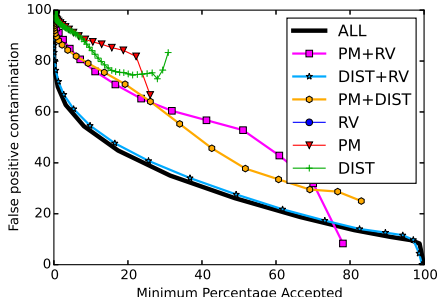} 
 \caption{In this figure we see the cost of lowering the membership probability threshold: the false positive contamination of the sample increases. The best way to reduce the false positive contamination is to obtain more of the 6 kinematic elements for the sample of stars, but even in this case, additional methods (spectroscopic confirmation, for example) will be needed to weed out the sample.}
   \label{fig:beta_pic_falsepositive}
\end{center}
\end{figure}

Another good reason to obtain more kinematic information is to reduce the false positive contamination (Fig.\,\ref{fig:beta_pic_falsepositive}) of a survey. With only proper motion and a fixed cutoff of 20\% probability, over 80\% of the objects in a survey will be contaminants (to be weeded out by some other technique, such as spectroscopy). Having complete data for the star of interest reduces the contamination rate to just above 40\%.

\subsection{\bf Not All Young Stars Belong to a Known Moving Group}
\begin{figure}[h]
\begin{center}
\includegraphics[width=3.4in]{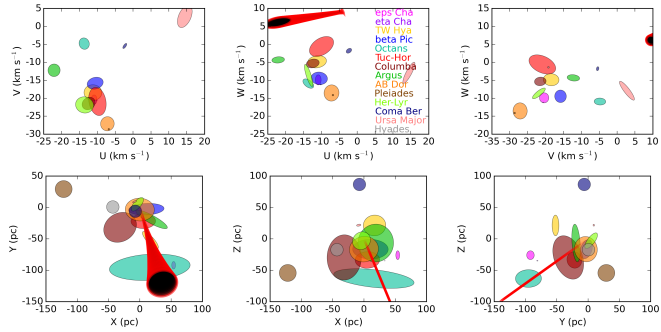} 
 \caption{2MASS 0447-7412 has a significant lithium equivalent width, 50$\pm$20 m\AA \cite[(Murphy \& Lawson 2015)]{Murphy2015}, a proper motion, a radial velocity, and no distance. If we generate a range of reasonable distances for the star, it could spatially be within the Octans (cyan) moving group, but it cannot possibly share the space velocity of Octans, and cannot be a member (as previously noted by Murphy)
   \label{fig:nongroup}}
\end{center}
\end{figure}

Virtually all lithium is primordial and easily fused by objects over roughly 60 Jupiter masses. Therefore, a low mass object with detectable lithium must either be a brown dwarf, or a young star. \cite[Riedel et al. (submitted)]{Riedel2015} assembled a sample of roughly 500 lithium-detected nearby stars and used those stars as a test sample for LACEwING, reasoning that these should be the best nearby young moving group members. However, nearly half the sample did not match any of the known nearby young moving groups LACEwING tests for. The same lithium-detected sample was run through other moving group codes - BANYAN I and II, and the Convergence code - with similar results. Some lithium-detected stars were not sorted into moving groups (Fig.\,\ref{fig:nongroup}).

There are three potential reasons for this: 1.) There are additional nearby young stellar groups not currently known or accounted for; 2.) there is a young field population not part of any previously identified groups; 3.) our current understanding of the kinematics of moving groups is incomplete. The fact that multiple moving group codes (each with a different algorithm and kinematic model of the nearby young moving groups) could not identify large numbers of stars as members of nearby young moving groups suggests that the explanation probably involves some combination of the first two points. Ultimately, the message is clear: Young stars do not need to be members of a known nearby young moving group. ``None of the above'' can be a correct answer.

\section{Conclusions}

As we enter the third decade of nearby young star research, there is much more to be done from the standpoint of membership statistics. Moving group identification codes such as LACEwING will continue to provide essential tools in studies of stellar and planetary formation and evolution, but care must still be taken in interpreting their results.

\begin{discussion}

\discuss{Shkolnik}{What about chemical abundance tagging?}

\discuss{Riedel}{LACEwING does only one thing: it uses kinematic and spatial data to evaluate potential memberships in moving groups. It should be used alongside spectroscopic methods to confirm that the star(s) in question is/are actually young, and have appropriate ages for the group LACEwING suggests is the most probable host.}

\discuss{Zuckerman}{Why isn't Carina-Near in the code?}

\discuss{Riedel}{There wasn't enough in the literature about it. The same goes for Cas-Tau, 32 Ori, and Alessi 13. But they can be added fairly easily - new ellipsoids, new simulation.}

\end{discussion}

\end{document}